# PURRS: Towards Computer Algebra Support for Fully Automatic Worst-Case Complexity Analysis[*]

Roberto Bagnara, Andrea Pescetti, Alessandro Zaccagnini, and Enea Zaffanella

Department of Mathematics, University of Parma, Italy
{bagnara,pescetti,zaffanella}@cs.unipr.it, zaccagni@math.unipr.it

**Abstract.** Fully automatic worst-case complexity analysis has a number of applications in computer-assisted program manipulation. A classical and powerful approach to complexity analysis consists in formally deriving, from the program syntax, a set of constraints expressing bounds on the resources required by the program, which are then solved, possibly applying safe approximations. In several interesting cases, these constraints take the form of recurrence relations. While techniques for solving recurrences are known and implemented in several computer algebra systems, these do not completely fulfill the needs of fully automatic complexity analysis: they only deal with a somewhat restricted class of recurrence relations, or sometimes require user intervention, or they are restricted to the computation of exact solutions that are often so complex to be unmanageable, and thus useless in practice. In this paper we briefly describe PURRS, a system and software library aimed at providing all the computer algebra services needed by applications performing or exploiting the results of worst-case complexity analyses. The capabilities of the system are illustrated by means of examples derived from the analysis of programs written in a domain-specific functional programming language for real-time embedded systems.

## 1 Introduction

Complexity analysis aims at the derivation of bounds to the complexity of algorithms, processes and data structures. In particular, the results of partly or wholly automated worst-case complexity analyses can be used, e.g., to decide whether mobile agents should be allowed to run in a given context, prove that the code for an embedded system satisfies the space and time constraints of the target hardware platform, assist programmers in reasoning about the behavior of programs, guide applications of optimized program transformations, and discover efficiency bugs that are otherwise very difficult to detect. Note that here we restrict attention to *worst-case complexity*: average-case complexity analysis is also a very interesting, but rather different research topic [9]. Moreover, we are interested in deriving proper upper and lower bounds, which are valid for all possible inputs, rather than asymptotic bounds.

---

[*] The work of R. Bagnara, A. Pescetti and E. Zaffanella has been partly supported by PRIN project "AIDA — Abstract Interpretation: Design and Applications." The work of A. Zaccagnini has been partly supported by PRIN project "Zeta and L Functions and Diophantine Problems in Number Theory."

Recurrence relations play an important role in the field of complexity analysis, since worst-case complexity measures of programs can often be very elegantly expressed by means of such relations. This is especially the case for the functional [3,4,14] and logic programming paradigms [6,7,8]. Therefore there is significant demand for efficient software systems capable, in a completely automatic way, of solving or approximating the solutions to systems of recurrence relations, yet providing results that are usable in practice. Up to now, this demand has been fulfilled only in a quite unsatisfactory way. In the systems described in [6,7,8] only recurrences belonging to a very restricted class are handled, essentially by pattern-matching.[1] The implementation of [3,4] is based on Mathematica, a remarkably powerful tool, which however is reported to sometimes exhaust all the available system memory in the attempt of finding closed-form exact solutions [3]. In these cases approximate solutions (in the form of upper and lower bounds for the exact solution) should be preferred, since their precision is sufficient for most practical purposes; moreover, often exact solutions are too complex to be useful. By resorting to approximations it is also possible to deal with classes of (generalized) recurrence relations that do not always admit closed-form solutions, such as the recurrences arising from the complexity analysis of *divide and conquer* algorithms. However, no computer algebra system we know of provides adequate support for the computation of upper or lower approximations of the solutions of recurrence relations.

The objective of the PURRS project (*Parma University's Recurrence Relation Solver*, see http://www.cs.unipr.it/purrs/) is the development of techniques and tools to provide all the computer-algebra services needed for efficiently computing the exact solution or manageable approximations of the solution of recurrence relations that arise when performing fully automated worst-case complexity analysis. A software library, also called PURRS, is actively being developed and is the subject of this paper. Space reasons do not allow to give more than a sketchy description of its features, but PURRS is free software released under the GNU General Public License: code and documentation can be downloaded at http://www.cs.unipr.it/purrs/. Concerning applications, PURRS is being integrated with a complexity analyzer (written by Pedro Vasconcelos, University of St Andrews, UK) aimed at deriving (possibly tight) upper bounds to the amount of stack and heap space consumed by programs written in *Hume* (http://www.hume-lang.org/), a functionally-inspired language for resource critical applications, including real-time embedded and safety-critical systems.

---

[1] These systems also use floating point numbers without controlled rounding so that correctness is, in general, compromised.

Several examples reported in the sequel come from the analysis of *Hume* programs: see also [14].

## 2 The PURRS Library

The PURRS library, which is written in C++, includes a number of mathematical tools that provide the functionalities required for both solving and approximating recurrence relations and to manipulate the results thus obtained. These tools include a solver for algebraic equations with rational coefficients and a sophisticate simplification apparatus that can handle, among other things, binomial coefficients and exponentials, as well as symbolic sums and products and including the Gosper's algorithm and a generalization of Zeilberger's algorithm [11,12] (the full implementation of all algorithms deriving from holonomy theory [12] is work in progress). PURRS also includes very efficient algorithms for proving statements of the form $\forall n \in \mathbb{N}\colon f(n) = 0$, where $f$ belongs to a quite large family of functions [1].

We will first sketch the techniques used to (approximately) solve recurrence relations in one argument. We will touch later the case of recurrences in more than one argument, often arising in a concrete complexity analysis, as well as the use of more aggressive approximation techniques. The solution process implemented by PURRS consists in an initial classification phase, where recurrences are categorized into one of five classes: these classes, which are characterized by different solution or approximation techniques, are briefly described in the following paragraphs.

*Linear recurrences of finite order with constant coefficients* (e.g., $x_n = 5x_{n-1} - 6x_{n-2} + n^2$) are very important for complexity analysis. Notice that, while single recurrences occurring in practice seldom have an order greater than 2, recurrences of higher order arise from transformation techniques mapping the resolution of a system of recurrences into the resolution of a single recurrence. The method employed in PURRS to solve these recurrences is an elaboration of the ideas proposed in [10] and [5], supplemented by *order-reduction* and other techniques that extend the class of recurrences that can be solved in a completely algorithmic way. All the details about the resolution of such recurrences are available in [2].

*Linear recurrences of finite order with variable coefficients* (e.g., $x_n = nx_{n-1} + 2$). While no general solution method is known, PURRS currently solves recurrences of this kind that are (possibly after the order-reduction step) of the first order. For higher-order recurrences, work is in progress to incorporate other methods (based, among other things, on Zeilberger's algorithm) that can be applied to find polynomial and *hypergeometric* solutions [12].

*Non-linear recurrences of finite order* (e.g., $x_n = 3x_{n-1}^2$) are known not to be generally solvable. PURRS handles some special cases by linearizing the recurrence using range transformations so as to obtain a recurrence belonging to one of the previous classes.

*Infinite order recurrences* (e.g., $x_n = n/2 + n \sum_{k=0}^{n-1} x_k$). Here $x_n$ depends on *all* previous values, and not only on a fixed number of them. Neither Mathematica nor other computer algebra systems we know of directly support this kind of recurrences. PURRS is able to transform and solve a class of such recurrences.

*Divide-and-conquer recurrences* (e.g., $x_n = 2x_{n/2} + n - 1$). As closed-form solutions may not exist for this kind of generalized recurrences, approximations are, in this case, unavoidable. PURRS, unlike any computer algebra system we are aware of, is able to derive upper and lower bounds for the solution, under the hypotheses described below, that are valid for each $n \in \mathbb{N} \setminus \{0\}$ for which the recurrence is well-defined. Recurrences of the general form $x_n = \alpha x_{n/\beta} + g(n)$ are approximated by PURRS when $\alpha$ and $\beta$ are rational numbers such that $\alpha > 0$, $\beta > 1$ and $g(n)$ is a non-negative, non-decreasing function. All these hypotheses are generally satisfied for the recurrences arising from worst-case complexity analysis. As an example of the capabilities of PURRS, from the analysis of Strassen's algorithm [13] we get $x_n = 7x_{n/2} + 9n^2/2$, with $x_1 = 1$, which is approximated by $n^{(\log 7)/\log 2} - \frac{3}{2}n^2 \leq x_n \leq 7n^{(\log 7)/\log 2} - 6n^2$. Another example comes from the analysis of the mergesort algorithm: $x_n = 2x_{n/2} + n - 1$ is approximated by PURRS with $h(n) - 3n + 3 + \frac{1}{2}nx_1 \leq x_n \leq h(n) - \frac{1}{2}n + 1 + nx_1$, where $h(n) = n(\log n)/\log 2$. Notice that PURRS has correctly determined the asymptotic formula $x_n \sim n(\log n)/\log 2$.

*Multivariate recurrences.* Up to now, we have only dealt with univariate recurrence relations. However, automatic complexity analysis frequently leads to the synthesis of multivariate recurrence relations. While multivariate recurrences are really hard to solve in general, in many cases they can be converted into univariate recurrences so that all the techniques presented above become applicable. PURRS is often able to automatically perform such a rewriting step. A very frequent and interesting case happens when the difference between the arguments of the unknown function $x$ is constant among all its occurrences in the multivariate recurrence relation. For instance, this happens for any recurrence relation having the form $x_{m,n} = f(x_{m-1,n-1})$, where the difference between the first and second argument of $x$ is always $m - n$. Such a recurrence can be rewritten as $y_t = f(y_{t-1})$, where $y_{t-k} = x_{m-k,n-k}$, for all $k \in \mathbb{N}$. Another interesting case, similar to the one above, is when the sum of the arguments of the unknown function $x$ is constant. For instance, multivariate recurrences of the form $x_{m,n} = f(x_{m+1,n-1})$ can be rewritten as $y_t = f(y_{t-1})$, where $y_{t-k} = x_{m+k,n-k}$, for all $k \in \mathbb{N}$. These rewriting techniques extend to recurrences involving more than two variables. As an example, the recurrence $x_{m,n} = a + x_{m-1,n+1}$ with initial conditions $x_{0,n} = 9$ results from the analysis of required stack depth for a *Hume* program for reversing a list using an accumulating parameter. PURRS can find the exact result $x_{m,n} = 9 + am$.

We are currently extending PURRS with approximation techniques that allow to determine simple yet precise upper and lower bounds to the solution of recurrences. Experience is suggesting that this is what is required by most of the applications in automatic, worst-case complexity analysis. These new approximation techniques can deal with all the recurrences defined above satisfying suitable non-negativity properties, and can also approximate recurrences that do *not* admit a closed-form solution. More specifically, without actually solving the recurrence, we determine functions $f, g_-, g_+ \colon \mathbb{N} \to \mathbb{R}$, all of the form $c \cdot n^d \cdot \alpha^n$ for suitable $c, d, \alpha \in \mathbb{R}$ where $d \geq 0$ and $\alpha > 0$, such that $f(n) - g_-(n) \leq x_n \leq f(n) + g_+(n)$ for all $n \in \mathbb{N}$. The method employed in the approximation depends on the detection of the asymptotically leading term of the solution, as described in a different context in [1]. For example, the recurrence $x_n = x_{n-1} + x_{n-3} + 2^n + n - 1$ arises from the manipulation of the system

$$\begin{cases} x_n = x_{n-1} + y_{n-1} + 2^n, \\ y_n = z_{n-1} + n - 1, \\ z_n = x_{n-1} + 1. \end{cases}$$

Assuming that the initial conditions $x_0$, $x_1$ and $x_2$ are all non negative, and letting $X = \max\{x_0, x_1, x_2\}$, we obtain

$$\frac{8}{3} 2^n - \frac{35}{3} \frac{\lambda}{\lambda - 1} \lambda^n \leq x_n \leq \frac{8}{3} 2^n + \lambda^n \frac{\lambda}{(\lambda - 1)^2}(X + 1),$$

where $\lambda$ is any upper bound for the positive root of $x^3 - x^2 - 1 = 0$.

*Acknowledgments* The authors are grateful to Tatiana Zolo who did so much for the development of PURRS and to Pedro Vasconcelos for the work done on the interface between PURRS and his analyzer and for the interesting and fruitful discussions we had on the subject of this paper and on related matters.